\documentclass[%
 reprint,
superscriptaddress,
 amsmath,amssymb,
 aps,
]{revtex4-2}

\usepackage{graphicx}
\usepackage{dcolumn}
\usepackage{bm}

\usepackage{upgreek}
\usepackage{physics}
\usepackage{braket}
\usepackage{xcolor}
\definecolor{BrewerSetRed}{RGB}{228,26,28}
\definecolor{BrewerSetBlue}{RGB}{55,126,184}
\definecolor{BrewerSetGreen}{RGB}{77,175,74}
\definecolor{BrewerSetPurple}{RGB}{152,78,163}
\definecolor{BrewerSetOrange}{RGB}{255,127,0}
\definecolor{BrewerSetYellow}{RGB}{236,176,23}
\definecolor{BrewerSetBrown}{RGB}{166,86,40}
\definecolor{BrewerSetPink}{RGB}{247,129,191}
\definecolor{BrewerSetGray}{RGB}{153,153,153}

\usepackage{hyperref}
\hypersetup{colorlinks,
    bookmarks,
	linkcolor={blue!75!black!80!yellow},
	citecolor={blue!75!black!80!yellow},
	urlcolor={blue!75!black!80!yellow}
}

\usepackage{tikz}
\usetikzlibrary{shapes.geometric,positioning}

\begin{document}

\preprint{APS/123-QED}

\title{Experimentally Probing Non-Hermitian Spectral Transition and Eigenstate Skewness}

\affiliation{Graduate Program in Acoustics, The Pennsylvania State University, University Park, PA 16802, USA}
\affiliation{Key Laboratory of Modern Acoustics and Institute of Acoustics, Nanjing University, Nanjing 210093, China}
\affiliation{Department of Physics, State Key Laboratory of Surface Physics, and Key Laboratory of Micro and Nano Photonic Structures (Ministry of Education), Fudan University, Shanghai, 200438, China}
\affiliation{NJU-Horizon Intelligent Audio Lab, Horizon Robotics, Beijing 100094, China}

\author{Jia-Xin Zhong}
\affiliation{Graduate Program in Acoustics, The Pennsylvania State University, University Park, PA 16802, USA}
\author{Jeewoo Kim}
\affiliation{Graduate Program in Acoustics, The Pennsylvania State University, University Park, PA 16802, USA}
\author{Kai Chen}
\affiliation{Key Laboratory of Modern Acoustics and Institute of Acoustics, Nanjing University, Nanjing 210093, China}
\affiliation{NJU-Horizon Intelligent Audio Lab, Horizon Robotics, Beijing 100094, China}
\author{Jing Lu}
\email{lujing@nju.edu.cn}
\affiliation{Key Laboratory of Modern Acoustics and Institute of Acoustics, Nanjing University, Nanjing 210093, China}
\affiliation{NJU-Horizon Intelligent Audio Lab, Horizon Robotics, Beijing 100094, China}
\author{Kun Ding}
\email{kunding@fudan.edu.cn}
\affiliation{Department of Physics, State Key Laboratory of Surface Physics, and Key Laboratory of Micro and Nano Photonic Structures (Ministry of Education), Fudan University, Shanghai, 200438, China}
\author{Yun Jing}
\email{yqj5201@psu.edu}
\affiliation{Graduate Program in Acoustics, The Pennsylvania State University, University Park, PA 16802, USA}

\begin{abstract}

Non-Hermitian (NH) systems exhibit intricate spectral topology arising from complex-valued eigenenergies, with positive/negative imaginary parts representing gain/loss.
Unlike the orthogonal eigenstates of Hermitian systems, NH systems feature left and right eigenstates that form a biorthogonal basis and can differ significantly, showcasing pronounced skewness between them.
These characteristics give rise to unique properties absent in Hermitian systems, such as the NH skin effect and ultra spectral sensitivity.
However, conventional experimental techniques are inadequate for directly measuring the complex-valued spectra and left and right eigenstates---key elements for enhancing our knowledge of NH physics. 
This challenge is particularly acute in higher-dimensional NH systems, where the spectra and eigenstates are highly sensitive to macroscopic shapes, lattice geometry, and boundary conditions, posing greater experimental demands compared to one-dimensional systems.
Here, we present a Green's function-based method that enables the direct measurement and characterization of both complex-valued energy spectra and the left and right eigenstates in arbitrary NH lattices. 
Using active acoustic crystals as the experimental platform, we observe spectral transitions and eigenstate skewness in two-dimensional NH lattices under both nonreciprocal and reciprocal conditions, with varied geometries and boundary conditions. 
Our approach renders complex spectral topology and left eigenstates experimentally accessible and practically meaningful, providing new insights into these quantities.
The results not only confirm recent theoretical predictions of higher-dimensional NH systems but also establish a universal and versatile framework for investigating complex spectral properties and NH dynamics across a wide range of physical platforms.
\end{abstract}

\pdfbookmark[1]{Title}{title} 

\maketitle

\section{Introduction}
\begin{figure*}[!htb]
\centering
\includegraphics[width=0.99\textwidth]{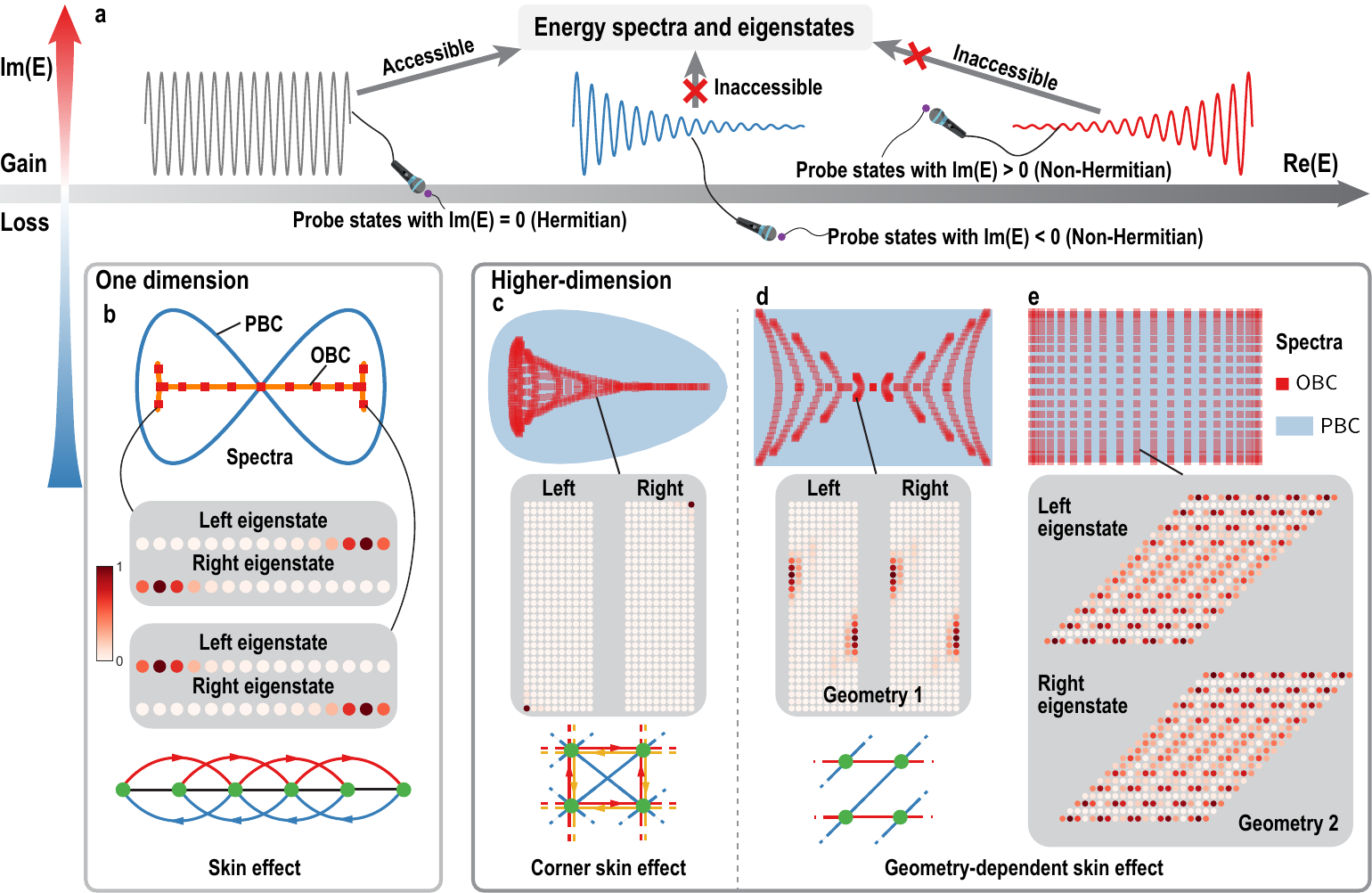}
\caption{
    \textbf{Challenges of probing energy spectra and eigenstates in NH lattices.}
    \textbf{a}, Schematic of probed time-domain signals when the lattice is excited by a source with a single-frequency sine signal.
    For a Hermitian system, where all states have zero imaginary part of their energies ($\Im(E)=0$), the source excitation produces stable signals, enabling reliable extraction of real-valued energy spectra and eigenstates.
    For a NH system, states with $\Im(E)<0$ exhibit signal decay due to loss, while those with $\Im(E)>0$ experience excessive amplification due to gain, challenging conventional experimental techniques and preventing access to complex-valued energy spectra and eigenstates.
    \textbf{b}, 1D NH lattice exhibiting the NHSE. 
    The panels from top to bottom display complex-valued energy spectra, skewed distributions of two representative eigenstates, and the schematic of the lattice structure.
    \textbf{c--e}, Higher-dimensional NH lattice exhibiting (\textbf{c}) corner NHSE and (\textbf{d, e}) geometry-dependent NHSE.
    The panels from top to bottom in \textbf{c, d} display spectra, eigenstates, and the lattice schematic.
    In the corner NHSE (\textbf{c}), the left and right eigenstates are highly skewed, while in geometry-dependent NHSE (\textbf{d, e}), the eigenstates are identical.
    The geometry-dependent NHSE appears in certain lattice configurations, such as rectangles (Geometry 1 in \textbf{d}), but vanishes in other shapes, such as parallelograms (Geometry 2 in \textbf{e}).
    Compared to 1D NH lattices, higher-dimensional systems exhibit more intricate spectral topology and eigenstate behaviors, amplifying experimental challenges.
    }
\label{fig:sketch}
\end{figure*}

Physical systems in the real world are inevitably open in various senses, making non-Hermitian (NH) physics more ubiquitous than its Hermitian counterpart. 
A hallmark of NH physics is the emergence of complex-valued eigenvalues and non-orthogonal eigenstates, which are absent in Hermitian systems  \cite{Kunst2018BiorthogonalBulkBoundaryCorrespondence, El-Ganainy2018NonHermitianPhysicsPT, Kawabata2019SymmetryTopologyNonHermitian, Ashida2020NonHermitianPhysics, Bergholtz2021ExceptionalTopologyNonHermitian, Ding2022NonHermitianTopologyExceptionalpoint}.
This paradigm shift towards NH systems has introduced novel concepts and phenomena, such as spectral topology \cite{Wang2021TopologicalComplexenergyBraiding, Wang2021GeneratingArbitraryTopological, Zhang2020CorrespondenceWindingNumbers, Zhang2021AcousticNonHermitianSkin, Ding2022NonHermitianTopologyExceptionalpoint}, the NH skin effect (NHSE) \cite{MartinezAlvarez2018NonHermitianRobustEdge, Yao2018EdgeStatesTopological, Kunst2018BiorthogonalBulkBoundaryCorrespondence, Yang2020NonHermitianBulkBoundaryCorrespondence, Xiao2020NonHermitianBulkBoundary,Wang2022NonHermitianMorphingTopological}, and exceptional points (EPs) \cite{Bender1998RealSpectraNonHermitian, Hodaei2017EnhancedSensitivityHigherorder, Tang2020ExceptionalNexusHybrid, Ding2022NonHermitianTopologyExceptionalpoint}, all of which are unattainable in Hermitian systems. 
While these effects have been well-studied in one-dimensional (1D) systems \cite{Kunst2018BiorthogonalBulkBoundaryCorrespondence, Yao2018EdgeStatesTopological, Yokomizo2019NonBlochBandTheory, Yang2020NonHermitianBulkBoundaryCorrespondence, Okuma2020TopologicalOriginNonHermitian, Zhang2020CorrespondenceWindingNumbers, Xue2021SimpleFormulasDirectional}, 
higher-dimensional NH systems exhibit substantially richer and more distinctive behaviors \cite{Yao2018NonHermitianChernBands, Liu2019SecondOrderTopologicalPhases, Yokomizo2023NonBlochBandsTwodimensional,  Wang2024AmoebaFormulationNonBloch, Hu2024TopologicalOriginNonHermitian, Luo2019HigherOrderTopologicalCorner, Zhang2022UniversalNonHermitianSkin,Franca2022NonHermitianPhysicsGain, Wan2023ObservationGeometrydependentSkin, Zhou2023ObservationGeometrydependentSkin, Wang2023ExperimentalRealizationGeometrydependenta,  Zhang2024EdgeTheoryNonHermitian, Shu2024UltraSpectralSensitivity, Song2024FragileNonBlochSpectrum}.
For example, as illustrated in Fig.~\ref{fig:sketch}, energy spectra and eigenstates in higher-dimensional NH systems display pronounced sensitivity to macroscopic shapes, lattice geometry, and impurities \cite{Luo2019HigherOrderTopologicalCorner,  Zhang2022UniversalNonHermitianSkin,Franca2022NonHermitianPhysicsGain, Wan2023ObservationGeometrydependentSkin, Zhou2023ObservationGeometrydependentSkin, Wang2023ExperimentalRealizationGeometrydependenta,  Zhang2024EdgeTheoryNonHermitian, Shu2024UltraSpectralSensitivity, Song2024FragileNonBlochSpectrum}.
These intriguing features have been predominately discussed in the theoretical framework; however, their physically measurable responses are crucial for investigating higher-dimensional NH systems, as real-energy observables ultimately hold the most significance.
This raises fundamental questions about the physical observability and detectability of the myst behaviors in higher-dimensional NH systems, underscoring the need for innovative experimental techniques to capture their complex properties and validate theoretical predictions.

In many physical systems, such as photonics, acoustics, and mechanics, eigenenergy spectra and eigenstates are typically investigated using pump-probe techniques, where the system is excited at specific sites and the frequency response is measured \cite{Susstrunk2015ObservationPhononicHelical, Brandenbourger2019NonreciprocalRoboticMetamaterials, Ma2019TopologicalPhasesAcoustic, Weidemann2020TopologicalFunnelingLight, Xiao2020NonHermitianBulkBoundary,Liu2021BulkDisclinationCorrespondence, Wu2022TopologicalPhononicsArising}. 
While effective for Hermitian systems, where real-valued eigenenergies yield well-defined frequency response peaks, these approaches fundamentally fail to capture the complexities of NH systems.
In NH systems, the presence of complex-valued eigenenergies, with imaginary components representing gain or loss, introduces complex-valued poles in the system's Green's functions.
Observing the eigenstates requires excitation near these poles, which is incompatible with conventional pump-probe techniques that rely on real frequency excitation (RFE) \cite{ Xiong2024TrackingIntrinsicNonHermitian}.
To address these limitations, previous efforts have introduced complex frequency excitation (CFE) methods, which allow the direct excitation of points in the complex frequency plane \cite{Li2020VirtualParityTimeSymmetry, Kim2022BoundsLightScattering, Kim2023LossCompensationSuperresolution, Guan2023OvercomingLossesSuperlenses, Guan2024CompensatingLossesPolariton, Zeng2024SynthesizedComplexfrequencyExcitation, Gu2022TransientNonHermitianSkin, Gao2024ControllingAcousticNonHermitian, Zhong2024HigherorderSkinEffect, Jiang2024ObservationNonHermitianBoundarya}.
This approach has been used to enhance the skin states in NH acoustic lattices and transmission line networks \cite{Gu2022TransientNonHermitianSkin, Gao2024ControllingAcousticNonHermitian, Zhong2024HigherorderSkinEffect, Jiang2024ObservationNonHermitianBoundarya}.
However, exciting lossy states in physical platforms using CFE presents its own challenges.
For instance, the CFE cannot target states located below others in the complex plane, as the effective gain of higher states overwhelms the excitation signal, leading to exponentially increased state distributions that obscure the target states (see Supplementary Material for more details). 
Existing experimental observations of the NHSE in low-loss NH systems are primarily based on the accumulation of massive boundary-localized skin states, serving only as signatures of right eigenstates, while left eigenstates have remained experimentally inaccessible.
This restriction prevents the direct investigation of state skewness, which captures the asymmetry between left and right eigenstates.
The above limitations are particularly pronounced in higher-dimensional NH systems, where the spectral topology and eigenstate behaviors are significantly more intricate compared to their 1D counterparts (Fig.~\ref{fig:sketch}).
Numerous recent theoretical predictions in this area have revealed fascinating and rich physics \cite{Yao2018NonHermitianChernBands, Liu2019SecondOrderTopologicalPhases, Yokomizo2023NonBlochBandsTwodimensional,  Wang2024AmoebaFormulationNonBloch, Hu2024TopologicalOriginNonHermitian, Luo2019HigherOrderTopologicalCorner, Zhang2022UniversalNonHermitianSkin,Franca2022NonHermitianPhysicsGain,  Zhang2024EdgeTheoryNonHermitian, Shu2024UltraSpectralSensitivity, Song2024FragileNonBlochSpectrum}, yet a comprehensive understanding of these phenomena necessitates direct measurements of complex-valued energy spectra and the left and right eigenstates.
This underscores the urgent need for methodologies capable of overcoming these experimental challenges.

\section{Green's function-based method for characterizing NH lattices}
\begin{figure*}[p]
\centering
\includegraphics[width=0.977\textwidth]{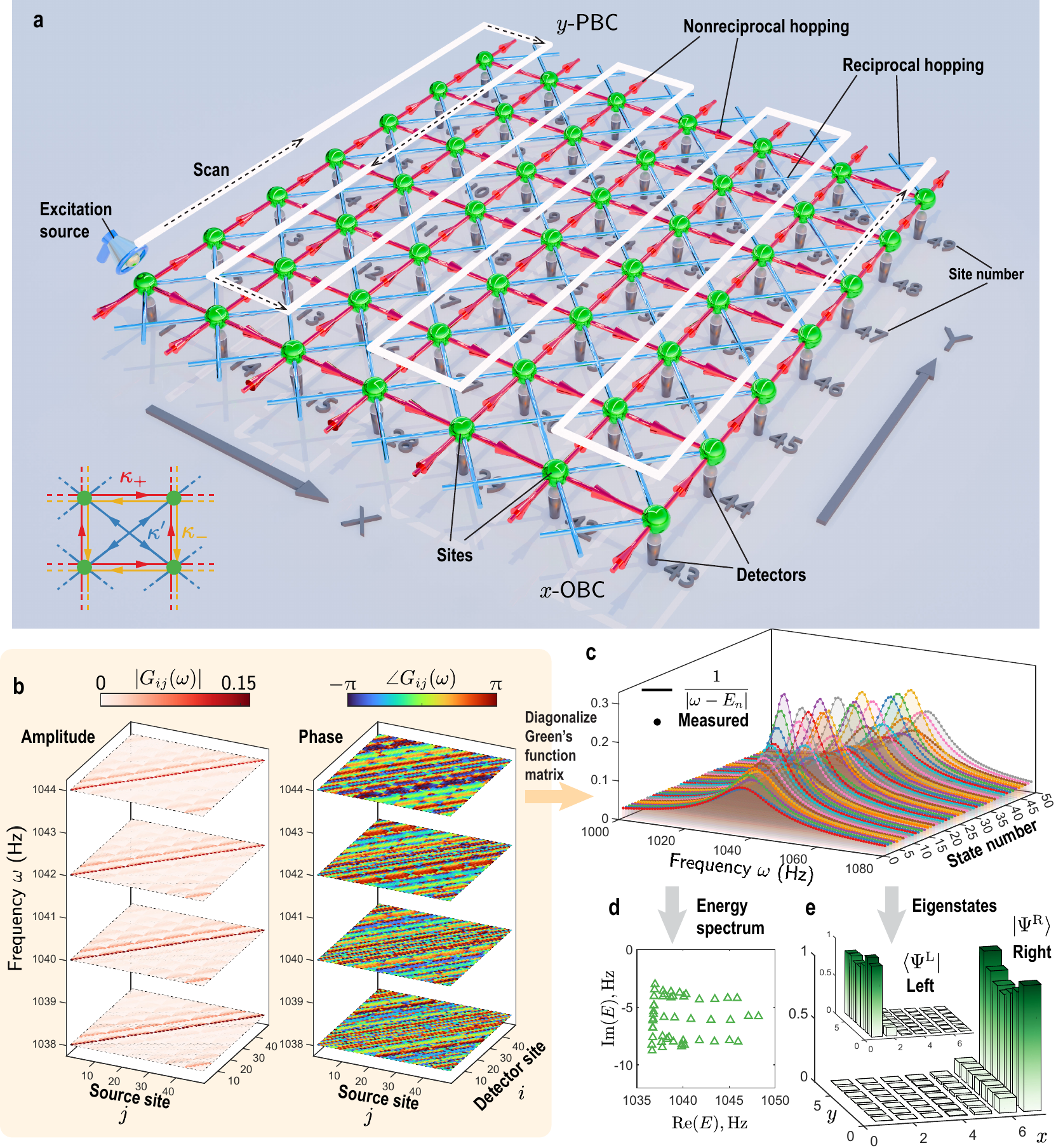}
\caption{
    \textbf{Illustration of the Green's function-based method for directly measuring the complex energy spectra and left and right eigenstates of NH systems.}
    \textbf{a}, Sketch of a $7\times 7$ nonreciprocal NH lattice with OBCs along the $x$-direction and PBCs along the $y$-direction (See Supplementary Materials for details on implementations of other boundary condition configurations).
    A source is sequentially excited along the white path, and the frequency responses of all sites are measured by detectors to obtain the full Green's function.
    The bottom-left inset provides a detailed schematic of the nonreciprocal NH lattice structure.
    The Bloch Hamiltonian of the system is $H_\mathrm{NR}(\vb{k}) = \omega_0+\kappa_+ \mathrm{e}^{-\mathrm{i}k_x} + \kappa_+ \mathrm{e}^{-\mathrm{i}k_y} - \kappa_- \mathrm{e}^{\mathrm{i}k_x} - \kappa_- \mathrm{e}^{\mathrm{i}k_y} + 4\kappa' \cos k_x \cos k_y  $ \cite{Wang2024AmoebaFormulationNonBloch}. 
    Parameters used in experiments are $\omega_0/(2\uppi) = 1040\,\mathrm{Hz} - 6\mathrm{i}\,\mathrm{Hz}, \kappa_+/(2\uppi)=2.72\,\mathrm{Hz}, \kappa_-/(2\uppi) = 0.48\,\mathrm{Hz}$ and $\kappa'/(2\uppi)=0.64\,\mathrm{Hz}$.
    \textbf{b}, Experimentally measured Green's functions of the lattice shown in $\textbf{a}$, showing (left) amplitude and (right) phase at four representative frequencies: 1038\,Hz, 1040\,Hz, 1042\,Hz, and 1044\,Hz.
    \textbf{c}, Magnitudes of diagonalized Green's functions obtained from the results in \textbf{b}.
(\textbf{d, e}) Experimental results: ($\textbf{d}$) energy spectrum and (\textbf{e}) left and right eigenstates derived from \textbf{c}.
}\label{fig:Green_fun}
\end{figure*}

To overcome the limitations of current experimental techniques in effectively investigating NH lattices, we present an experimental approach that allows for the direct measurement of both complex energy spectra and the left and right eigenstates in arbitrary NH lattices, irrespective of their dimensions, boundary conditions, macroscopic shapes, or lattice geometries.
To illustrate this method, we begin with a 2D single-band nonreciprocal lattice, as illustrated in Fig.~\ref{fig:Green_fun}a.
All lattice hoppings are experimentally implemented using loudspeaker-microphone pairs.
Our active acoustic lattice platform further allows for the realization of open, periodic, or mixed boundary conditions, as well as various lattice geometries (see Supplementary Materials for details).

In conventional experimental techniques, the system is excited at a single source position, and the magnitude response measured at multiple detector positions is used to represent the excited state.
However, these methods often neglect the phase response, limiting the completeness of the characterization.
In contrast, our method extracts both the \emph{magnitude and phase} of the frequency response, enabling the full characterization of the Green's function $G_{ij}(\omega)$, where $\omega$ is the excitation frequency, and $i$ and $j$ denote the detector (probe) and source (pump) positions, respectively.
In our experimental platform, the phase of the Green's function is obtained based on the measured acoustic pressure, with the excitation signal serving as the reference. 
Additionally, we systematically measure the Green's function for \emph{all possible combinations} of source and detector positions across the entire lattice (Figs.~\ref{fig:Green_fun}a--b).
Specifically, for a lattice with $L$ sites ($L=49$ in Fig.~\ref{fig:Green_fun}), we excite the system sequentially at each $j$-th site ($j=1,2,...,L$) using an acoustic source (loudspeaker) and measure the frequency response at every $i$-th site ($i=1,2,...,L$) using a detector (microphone), constructing the full Green's function matrix $G(\omega)$.
This process generates $L^2$ (2,401 in Fig.~\ref{fig:Green_fun}) Green's functions for this lattice configuration, making manual measurements impractical.
To overcome this challenge, we employ a custom-programmed data acquisition system to efficiently scan and record all results (see Supplementary Materials for details). 
By mapping all source-detector combinations, our approach offers a comprehensive characterization of the lattice response, surpassing traditional methods that typically measure responses excited by a source at a single site.

The Green's function matrix $G(\omega)$ is directly related to the lattice Hamiltonian $H$ via $G(\omega) = (\omega-H)^{-1}$.
Using the spectral decomposition, $H= \sum_n E_n \ket{\psi_n^\mathrm{R}} \bra{\psi_n^\mathrm{L}}$, the Green's function can be expressed as $G (\omega) = \sum_n \frac{1}{\omega-E_n}  \ket{\psi_n^\mathrm{R}}\bra{\psi_n^\mathrm{L}}$ \cite{Ashida2020NonHermitianPhysics, Ding2022NonHermitianTopologyExceptionalpoint}.
Here, $n$ is the state index, $E_n$ is the eigenvalue, and $\bra{\psi_n^\mathrm{L}}$ and $\ket{\psi_n^\mathrm{R}}$ are the binormalized left and right eigenstates of $H$, respectively.
When $G(\omega)$ acts on the right eigenstate $\ket{\psi_n^\mathrm{R}}$ from the right or on the left eigenstate $\bra{\psi_n^\mathrm{L}}$ from the left, it yields 
\begin{equation}
G(\omega) \ket{\psi_n^\mathrm{R}}    
=  
\frac{1}{\omega-E_n} 
 \ket{\psi_n^\mathrm{R}}    
 \qc 
 \bra{\psi_n^\mathrm{L}}  G(\omega)
=  
\frac{1}{\omega-E_n} 
 \bra{\psi_n^\mathrm{L}}
 .
\label{eq:green_spec}
\end{equation}
Equation~(\ref{eq:green_spec}) demonstrates that the left and right eigenstates of $G(\omega)$ are identical to those of $H$, and the eigenvalues of $G(\omega)$ are directly related to the Hamiltonian's eigenvalues $E_n$ by ${1}/({\omega-E_n})$.

By diagonalizing the experimentally measured Green's function matrix $G(\omega)$, we can directly extract both the complex energy spectrum $E_n$ and the left ($\bra{\psi_n^\mathrm{L}}$) and right ($\ket{\psi_n^\mathrm{R}}$) eigenstates of a NH lattice.
Figure~\ref{fig:Green_fun}c illustrates the eigenvalues of $G(\omega)$ at different excitation frequencies $\omega$.
These eigenvalues form $L=49$ well-defined single-peak curves that correspond to the magnitude of $1/(\omega-E_n)$.
By fitting each curve with this simple expression, we precisely determine both the real and imaginary parts of $E_n$, thereby obtaining the complete complex energy spectrum (Fig.~\ref{fig:Green_fun}d).
In addition to the spectrum, the left and right eigenstates can also be directly extracted as the eigenstates of $G(\omega)$.
The sum of all eigenstates, defined as
\begin{equation}
{\Psi^\mathrm{L}} \equiv \frac{1}{L}\sum_n {\abs{\psi_n^\mathrm{L}}^2}\qc
{\Psi^\mathrm{R}} \equiv \frac{1}{L}\sum_n {\abs{\psi_n^\mathrm{R}}^2},
\end{equation}
are depicted in Fig.~\ref{fig:Green_fun}e.
Our experimental results presented in Fig.~\ref{fig:Green_fun}d--e exhibit great agreement with the theoretical predictions provided in Supplementary Figure S4.
It is observed from Fig.~\ref{fig:Green_fun}d that the energy spectrum in higher-dimensional NH lattices forms a finite-size area in the complex plane, contrasting with the line-like spectra observed in 1D counterparts (Fig.~\ref{fig:sketch}).
Furthermore, Fig.~\ref{fig:Green_fun}e confirms the pronounced state skewness due to nonreciprocity, a feature not previously demonstrated in wave-based experimental platforms. 
To quantify this state skewness, we define a parameter 
\begin{equation}
    \gamma \equiv 1-\frac{1}{L}\sum_n\frac{\abs{\braket{\psi_n^\mathrm{L}\,|\,\psi_n^\mathrm{R}}}}{\sqrt{
        \braket{\psi_n^\mathrm{L}\,|\, \psi_n^\mathrm{L}}
        \braket{\psi_n^\mathrm{R}\,|\, \psi_n^\mathrm{R}}}}    ,
\end{equation}
where the second term represents the phase rigidity \cite{Wiersig2023PetermannFactorsPhase}.
The skewness parameter satisfies $0\leq \gamma \leq 1$, with larger values indicating higher state skewness. 
From the measured eigenstates in Fig.~\ref{fig:Green_fun}e, the state skewness is calculated as $\gamma_\mathrm{exp} = 0.79$, highlight the pronounced state skewness induced by the system nonreciprocity.

It is worth noting that this approach is valid as long as the system is free from higher-order poles in the Green's functions and eigenstate coalescence.
However, small disturbances such as background noise or system imperfections often naturally fulfill this condition, making the method broadly applicable.
The accuracy of the proposed method depends on relatively high signal-to-noise ratio (SNR) measurements of the Green's functions and precise phase measurements. 
At low SNR levels or with significant phase errors, the accuracy of the retrieved energy spectra and eigenstates deteriorates (see Supplementary Materials for a detailed robustness analysis).
In our experiments, the SNR is more than 40\,dB, and the phase error remains below $1^\circ$ in all cases, ensuring negligible impact on the measured results (see Supplementary Materials). 
Our method is inherently universal, as it relies on the Green's function formalism---a framework applicable to all wave-based physical systems. 
This universality broadens the method's applicability to diverse platforms, such as photonics, acoustics, and mechanics \cite{Brandenbourger2019NonreciprocalRoboticMetamaterials, Weidemann2020TopologicalFunnelingLight, Xiao2020NonHermitianBulkBoundary, Zhang2021AcousticNonHermitianSkin, Wang2022NonHermitianMorphingTopological, Wang2023NonHermitianTopologyStatic}.
By enabling direct measurement of complex energy spectra and both left and right eigenstates, our established method opens new avenues for exploring the intricate physics of higher-dimensional NH systems, particularly their dependence on geometric configurations, which were previously inaccessible with conventional experimental techniques.

\section{Nonreciprocal NH lattices}
\begin{figure*}[!htb]
\centering
\includegraphics[width=0.99\textwidth]{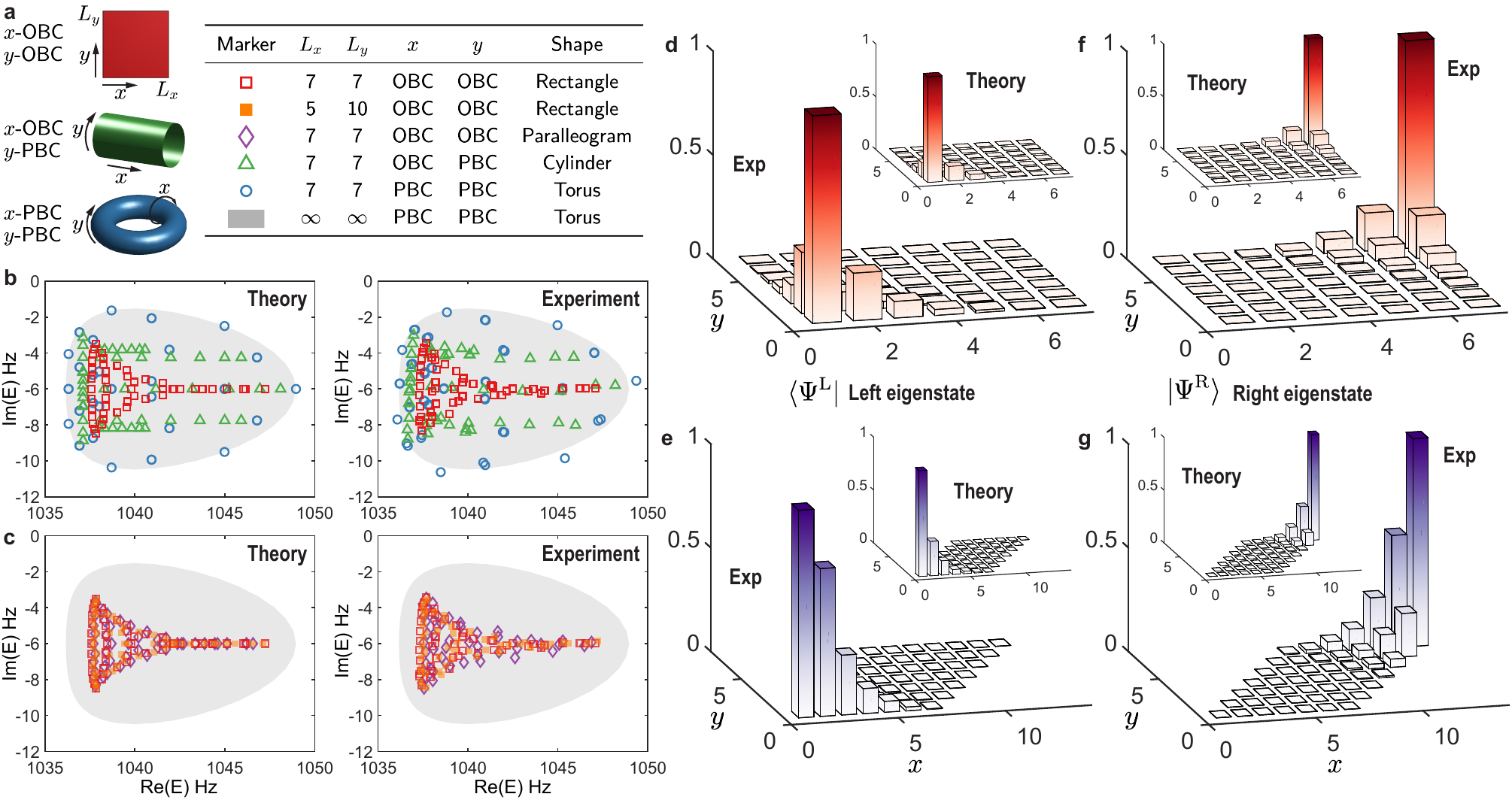}
\caption{\textbf{Experimental observation of spectral shrink and state skewness in nonreciprocal NH lattices.}
    Parameters used in experiments are $\omega_0/(2\uppi) = 1040\,\mathrm{Hz} - 6\mathrm{i}\,\mathrm{Hz}, \kappa_+/(2\uppi)=2.72\,\mathrm{Hz}, \kappa_-/(2\uppi) = 0.48\,\mathrm{Hz}$ and $\kappa'/(2\uppi)=0.64\,\mathrm{Hz}$.
\textbf{a}, Schematic representation of the lattice structure under different boundary conditions.
    (\textbf{b, c}) Complex energy spectra: Left panel, theoretical predictions using the tight-binding model; right panel, experimental results.
    The gray shaded region represents the thermodynamic limit with PBCs in both $x$ and $y$ directions.
    \textbf{b}, Comparison of energy spectra for different boundary conditions with the lattice size of $L_x = L_y=7$.
    \textbf{c}, Comparison of energy spectra for OBCs in different geometries.
    (\textbf{d, e}) Left and (\textbf{f, g}) right eigenstates, comparing (\textbf{d, f}) rectangular and (\textbf{e, g}) parallelogram lattices ($L_x=L_y=7$). 
}
\label{fig:zhongwang}
\end{figure*}

We apply our method to expand the results presented in Fig.~\ref{fig:Green_fun} by considering the same lattice but with various boundary conditions and lattice geometries (Fig.~\ref{fig:zhongwang}a).
Figure~\ref{fig:zhongwang}b experimentally verifies that the spectra of the 2D nonreciprocal NH lattice occupy a finite-size area in the complex plane, regardless of boundary conditions.
This observation marks a significant departure from the line-like spectra characteristic of 1D systems (Fig.~\ref{fig:sketch}).
Additionally, Fig.~\ref{fig:zhongwang}b reveals a hierarchical spectral relationship: the spectrum under full OBCs (both $x$-OBC and $y$-OBC, where OBC denotes open boundary condition) is fully contained within the spectrum under mixed boundary conditions ($x$-OBC and $y$-PBC, where PBC denotes periodic boundary conditions). 
This, in turn, is contained within the spectrum under full PBCs ($x$-PBC and $y$-PBC).
This spectral relationship mirrors that of 1D NH lattice, where the spectrum under OBCs is contained within that under semi-infinite boundary conditions \cite{Okuma2020TopologicalOriginNonHermitian}.
To investigate the spectral sensitivity to lattice geometry, which is a distinctive feature of higher-dimensional systems, Fig.~\ref{fig:zhongwang}c compares the energy spectra for lattices with different macroscopic shapes (rectangular and parallelogram) and aspect ratios, both under full OBCs.
Notably, the spectra consistently occupy similar finite-sized areas in the complex plane, demonstrating that the spectral topology is largely insensitive to lattice geometry.
These results represent the first experimental observation of such spectral relationships in higher-dimensional NH systems, advancing our understanding of their spectral topology and enriching the foundation for future investigations.

Figures~\ref{fig:zhongwang}d--g illustrate the measured left and right eigenstates under full OBCs for two representative lattice configurations (additional results are provided in Supplementary Materials).
The right eigenstates (Figs.~\ref{fig:zhongwang}f, g) are found to be strongly localized at the top-right corner of the lattice, a direct signature of the NHSE.
In contrast, the left eigenstates (Figs.~\ref{fig:zhongwang}d, e) exhibit localization at the opposite (bottom-left) corner.
In the configurations shown in Fig.~\ref{fig:zhongwang}d, f (Fig.~\ref{fig:zhongwang}e, g), the theoretical skewness value is $\gamma_\mathrm{theo} = 0.97$ ($\gamma_\mathrm{theo} = 0.99$), while the experimental results yield $\gamma_\mathrm{exp} = 0.90$ ($\gamma_\mathrm{exp} = 0.97$).
These results reveal substantial skewness between the left and right eigenstates, underscoring the intrinsic asymmetry in nonreciprocal systems.

Collectively, these findings demonstrate that the NHSE persists across different macroscopic lattice geometries, including rectangular and parallelogram configurations, highlighting the robustness of both spectral topology and state distributions in higher-dimensional nonreciprocal systems. 
The observed strong skewness between left and right eigenstates further confirms the inherent asymmetry of nonreciprocal NH systems.
Interestingly, higher-dimensional NH physics reveals that the NHSE, which in 1D systems requires nonreciprocity, can emerge even in reciprocal systems in higher dimensions (Fig.~\ref{fig:sketch}) \cite{Zhang2022UniversalNonHermitianSkin}. 
Moreover, higher-dimensional reciprocal NH systems exhibit strong sensitivity of energy spectra and eigenstate distributions to lattice geometry, adding richness to their physical behaviors. 
These unique properties underscore the necessity of investigating the spectral topology and state skewness in reciprocal NH systems.

\section{Reciprocal NH lattices}
\begin{figure*}[p]
\centering
\includegraphics[width=0.99\textwidth]{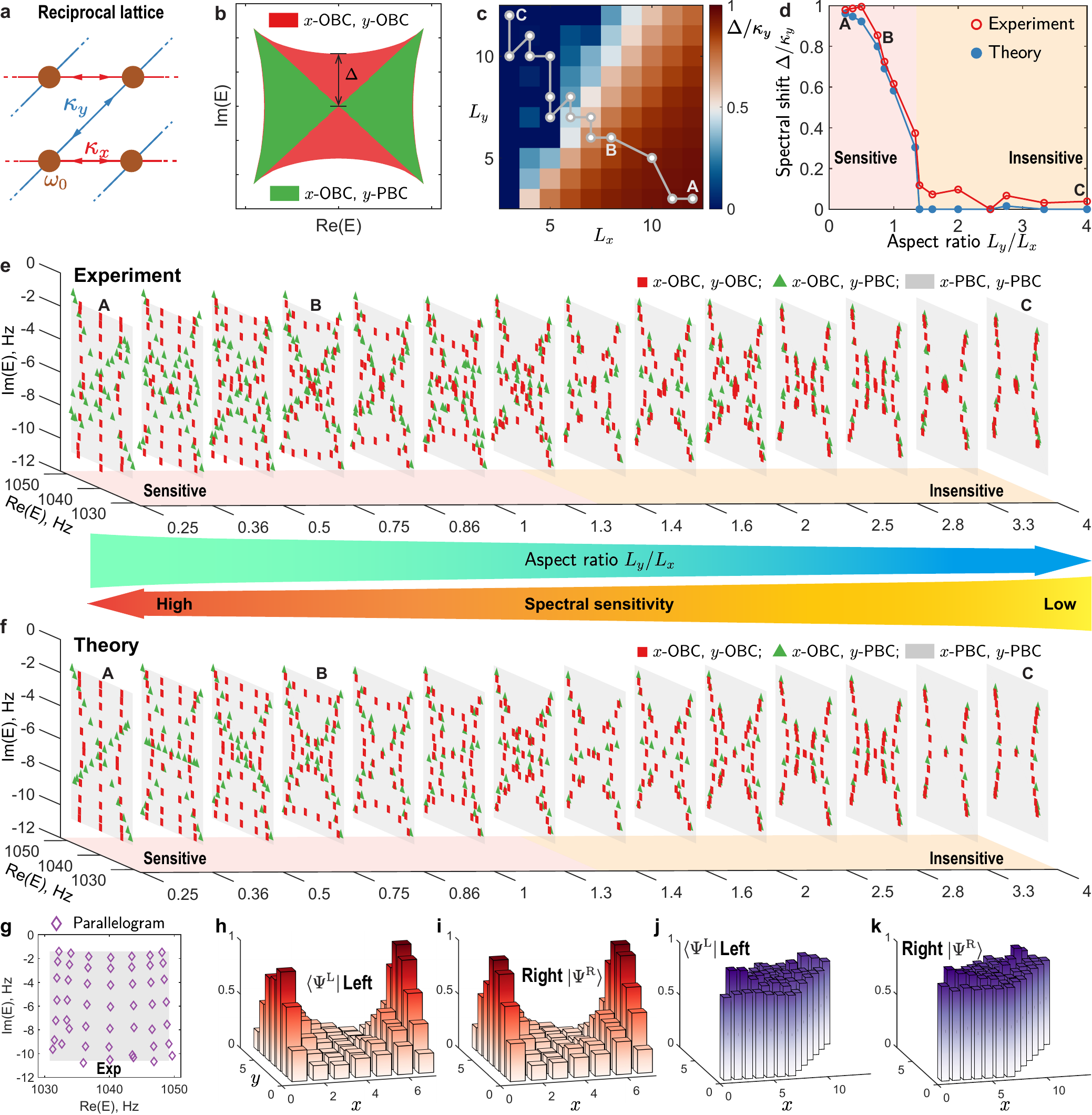}
\caption{
    \textbf{Experimental observation of spectral sensitivity transitions in reciprocal NH lattices.}
    \textbf{a}, Schematic representation of the lattice structure.
    The Bloch Hamiltonian of this system is $H_\mathrm{R}(\vb{k}) =\omega_0+ 2 \kappa_x \cos k_x+ 2 \kappa_y \cos(k_x+k_y) $ \cite{Zhang2024EdgeTheoryNonHermitian}.
    Parameters used in experiments are $\omega_0/(2\uppi) = 1040\,\mathrm{Hz} - 6\mathrm{i}\,\mathrm{Hz}, \kappa_x/(2\uppi)=2.25\mathrm{i}\,\mathrm{Hz}$ and $\kappa_y/(2\uppi)=4.5\,\mathrm{Hz}$.
    \textbf{b}, Definition of the spectral shift $\Delta$, quantifying the deviation of the spectrum of the rectangular lattice under full OBCs.
    \textbf{c}, Sensitivity diagram from the tight-binding model for various lattices sizes.
    \textbf{d}, Experimental results and theoretical predictions of the spectral shift as a function of selected aspect ratio configurations, also annotated in \textbf{c}.
    \textbf{e, f} Experimental results and theoretical predictions of energy spectra for different aspect ratios, showing transitions in spectral topology.
    (\textbf{g}) Energy spectra of parallelogram lattices ($L_x=L_y=7$) obtained experimentally.
    \textbf{g--k} Experimental results of (\textbf{h, j}) left and (\textbf{i, k}) right eigenstates, comparing (\textbf{h, i}) rectangular and (\textbf{j, k}) parallelogram lattices ($L_x=L_y=7$), demonstrating geometry-dependent spectral and eigenstate behaviors.
}
\label{fig:kaisun}
\end{figure*}

We next investigate higher-dimensional reciprocal NH lattices, whose structure is illustrated in Fig.~\ref{fig:kaisun}a.
Unlike their nonreciprocal counterparts, the energy spectra of reciprocal systems exhibit pronounced sensitivity to lattice geometry \cite{Zhang2022UniversalNonHermitianSkin, Shu2024UltraSpectralSensitivity}. 
To quantify this sensitivity, we adopt the previously defined parameter, spectral shift $\Delta$, which measures the deviation between the spectra of rectangular lattices under full OBCs and mixed boundary conditions (Fig.~\ref{fig:kaisun}b) \cite{Shu2024UltraSpectralSensitivity}.
The theoretical sensitivity diagram in Fig.~\ref{fig:kaisun}c reveals a sharp decrease in $\Delta$ when the aspect ratio $\alpha\equiv L_y/L_x$ exceeds a critical threshold $\alpha_\mathrm{c} \approx 1.35$. 
This trend is verified by our experimental results at selected aspect ratio configurations (Figs.~\ref{fig:kaisun}d--f).
For instance, for \emph{insensitive} lattice (labelled `C' in Figs.~\ref{fig:kaisun}d--f) with $\alpha = 4 > \alpha_\mathrm{c}$, the spectral area under full OBCs is entirely contained within the spectral area under mixed boundary conditions.
The experimentally measured spectral shift is $\Delta_\mathrm{exp}/\kappa_y=0.039$ closely matching the theoretical value of $\Delta_\mathrm{theo}/\kappa_y =0$.
In contrast, in \emph{sensitive} configurations where $\alpha < \alpha_\mathrm{c}$, the spectral topology undergoes significant changes: portions of the spectral area under mixed boundary conditions extend beyond those under full OBCs.
This behavior diverges from the spectral area relationship observed in nonreciprocal NH lattices.
For two representative configurations, labelled `A' or `B' in Figs.~\ref{fig:kaisun}d--f, the experimentally measured spectral shifts are $\Delta_\mathrm{exp} /\kappa_y = 0.98$ and $\Delta_\mathrm{exp}/\kappa_y = 0.99$, respectively.
Further insights into spectral behavior are revealed when comparing lattices with varying macroscopic geometries.
As shown in Fig.~\ref{fig:kaisun}g, the energy spectrum of a parallelogram lattice under full OBCs closely resembles that of the system under full PBCs. 
This contrasts with nonreciprocal systems, where the spectral area under full OBCs consistently occupies only a portion of the PBC spectrum, irrespective of lattice geometry (Fig.~\ref{fig:zhongwang}d). 
These findings experimentally confirmed the unique sensitivity of reciprocal higher-dimensional NH systems to geometric factors \cite{Shu2024UltraSpectralSensitivity}.
Our results highlight previously unobserved spectral topologies that are absent in both 1D and higher-dimensional nonreciprocal NH lattices, marking a significant advance in understanding the interplay between geometry and spectral topology in NH physics.

The eigenstate distributions in higher-dimensional reciprocal NH systems also exhibit distinctive characteristics.
Figures~\ref{fig:kaisun}h--k compare the left and right eigenstates for rectangular and parallelogram lattices.
For rectangular lattices, the eigenstates exhibit boundary-localized skin states.
In contrast, for parallelogram lattices, the eigenstates remain bulk-like, showing no localization.
This dependence of localization on the macroscopic lattice geometry demonstrates that the NHSE in reciprocal systems is sensitive to lattice shape, a feature absent in nonreciprocal lattices where localization is robust against geometric changes (Fig.~\ref{fig:zhongwang}d).
Additionally, the left and right eigenstates in reciprocal systems show minimal skewness, with  $\gamma_\mathrm{exp} = 0.002$ and 0.0015 for the configurations in Figs.~\ref{fig:kaisun}h--i and \ref{fig:kaisun}j--k, respectively.
Their nearly identical spatial distributions, regardless of lattice geometry, highlight the symmetric nature of reciprocal systems.
This behavior contrasts with nonreciprocal systems, where significant skewness arises between left and right eigenstates, as demonstrated in Figs.~\ref{fig:zhongwang}d--g.

\section{Conclusion and discussion}
In this work, we developed a universal Green's function-based method to directly measure the complex energy spectra and both the left and right eigenstates of higher-dimensional NH lattices.
This method addresses the limitations of conventional experimental techniques, which are typically restricted to measuring partial system responses and fail to capture the full complexity of NH systems.
Using this approach, we observed spectral transition and eigenstate skewness in 2D acoustic crystals under both nonreciprocal and reciprocal conditions, significantly enhancing our understanding of higher-dimensional NH physics.
Our approach is inherently universal, as it relies on the Green's function formalism, a framework applicable to all wave-based physical systems.
This universality enables seamless extensions to other platforms, including photonics and mechanics \cite{Brandenbourger2019NonreciprocalRoboticMetamaterials, Ma2019TopologicalPhasesAcoustic, Weidemann2020TopologicalFunnelingLight, Xiao2020NonHermitianBulkBoundary, Liu2021BulkDisclinationCorrespondence, Xue2022TopologicalAcoustics, Wu2022TopologicalPhononicsArising, Lu2023NonHermitianTopologicalPhononic, Wang2023NonHermitianTopologyStatic}.
Looking forward, our proposed method provides a powerful experimental tool to validate a wide range of advanced theoretical predictions in NH physics. 
These include spectral graph topology \cite{Tai2023ZoologyNonHermitianSpectra}, spectral density of states \cite{Wang2024AmoebaFormulationNonBloch}, defect-induced states \cite{Bhargava2021NonHermitianSkinEffect, Schindler2021DislocationNonHermitianSkin, Panigrahi2022NonHermitianDislocationModes, Sun2021GeometricResponseDisclinationInduced, Banerjee2024TopologicalDisclinationStatesa}, inner NHSE in fractal lattices \cite{Manna2023InnerSkinEffects}, critical NHSE \cite{Li2020CriticalNonHermitianSkin}, and observation of non-Euclidean NH lattices \cite{Hu2024UnveilingNonHermitianSpectral}. 
All of these phenomena remain experimentally elusive due to the inherent limitations of existing techniques. 
By enabling the experimental realization and investigation of these phenomena, our method not only bridges the gap between theory and experiment but also opens avenues for uncovering new and richer physics, potentially leading to more discoveries in the field of NH systems.


\begin{acknowledgments}
Y. J. thanks the support of startup
funds from Penn State University and NSF awards
2039463 and 195122. 
K. D. thanks the support of the National Key R\&D Program of China (No. 2022YFA1404500, No. 2022YFA1404701), the National Natural Science Foundation of China (No. 12174072, No. 2021hwyq05).
J. L. thanks the support of the National Natural Science Foundation of China (No. 12274221).
\end{acknowledgments}

\bibliography{bibtex}

\end{document}